# A Geography of Participation in IT-Mediated Crowds

John Prpić
Beedie School of Business,
Simon Fraser University
prpic@sfu.ca

Prashant Shukla
Beedie School of Business,
Simon Fraser University
Rotman School of Management,
University of Toronto
pshukla@sfu.ca

Yannig Roth
Université Paris 1 Panthéon
Sorbonne (PRISM)
yannigroth@gmail.com

Jean-François Lemoine
ESSCA Ecole de Management
Université Paris 1 Panthéon
Sorbonne (PRISM)
jflemoine30@hotmail.com

## Abstract

In this work we seek to understand how differences in location effect participation outcomes in IT-mediated crowds. To do so, we operationalize Crowd Capital Theory with data from a popular international creative crowdsourcing site, to determine whether regional differences exist in crowdsourcing participation outcomes. We present the results of our investigation from data encompassing 1,858,202 observations from 28,214 crowd members on 94 different projects in 2012. Using probit regressions to isolate geographic effects by continental region, we find significant variation across regions in crowdsourcing participation. In doing so, we contribute to the literature by illustrating that geography matters in respect to crowd participation. Further, our work illustrates an initial validation of Crowd Capital Theory as a useful theoretical model to guide empirical inquiry in the fast-growing domain of IT-mediated crowds.

## 1. Introduction

Crowdsourcing, a term popularized by Wired magazine contributor Jeff Howe [1, 2], involves organizations using IT to engage crowds of individuals for the purposes of completing tasks, solving problems, or generating ideas. In the last decade, many organizations have turned to crowdsourcing to engage with consumers, accelerate their innovation cycles, and to find new ideas for their brands [3, 4, 5]. As crowdsourcing has become an increasingly common method for organizations to gather IT-mediated input from individuals, the importance of understanding the nature of crowds has similarly increased. While many sing the praises of the global distribution of crowd participants at crowdsourcing platforms like eYeka, TopCoder, and Tongal, research has yet to emerge that empirically investigates the role of crowd member geography on crowdsourcing participation.

To achieve these aims we bound our investigation using Crowd Capital Theory (CCT) as our theoretical framework, which draws upon the knowledge-based view of organizations to explain how and why IT-mediated crowds can generate value for organizations [6]. In this study, we operationalize CCT with data provided to us by a leading global crowdsourcing platform which hosts crowdsourcing contests on behalf of major corporations. These contests are visible to anyone with internet access, and participation is free and open to all internet users. Therefore, the empirical context is very well-suited for our inquiry into the geography of Crowd Capital participation.

In the ensuing sections of this work we undertake this research design by first reviewing the crowdsourcing literature to establish the empirical context. Then, we introduce Crowd Capital Theory (CCT), establishing the theoretical motivation for our investigation, bounding our hypothesis, and structuring our data operationalizations. From here we introduce our research methods, detailing the data collection and data analysis undertaken, before establishing our findings. We then discuss the ramifications of our findings for both the research and practitioner communities, before concluding with a set of new research questions emanating directly from our work.

## 2. Crowdsourcing

Crowdsourcing is an IT-mediated problem solving, idea-generation, and production model that leverages distributed intelligence [7]. Problem-solving, idea-generation and production can be "crowd-sourced" by different means [8] such as micro-tasking (asking individuals to execute short tasks online for pay at virtual labor markets), open collaboration (asking individuals to volunteer contributions online) or through tournament-based competitions (where individuals submit contributions with the hope of winning a prize). The latter approach is increasingly used in the fields of innovation and marketing, as companies are posting their challenges through broadcast search on crowdsourcing platforms like eYeka, InnoCentive, Kaggle, Tongal or Topcoder, to access the crowds of motivated and skilled participants that have coalesced at these web properties.

### 2.1 Crowdsourcing in Use

Even though crowdsourcing is a very recent phenomenon, the process has been used in a variety of contexts. Companies can launch crowdsourcing initiatives on their own branded platforms, like Dell did with IdeaStorm, or they can commission crowdsourcing intermediaries to host contests on their web properties for a fee [4].

One way to engage in crowdsourcing initiatives for a company is to start with a private online platform. In order to benefit from the creativity of the crowd, companies create branded platforms, such as Dell's IdeaStorm, Starbucks' myStarbucksIdea, or Nokia's IdeasProject, where individuals can contribute ideas and suggestions, allowing the organization to benefit from the scale and diversity of the crowd to gather innovative and consumer-rooted ideas [4, 9, 10,11].

Another method by which to initiate crowdsourcing is to externalize the entire process to an intermediary whose job it is to organize crowdsourcing, often in the form of a contest, on behalf of companies [12, 13]. These intermediaries are specific in that they leverage a private community of contributors who participate in contests sponsored by client organizations [9, 14]. When applied to innovation, these intermediaries are usually called "innomediaries" [11] or idea marketplaces [15]. These intermediaries count on self-selected crowd contributions for the supply and/or selection of ideas and designs [5, 9].

### 3. Crowd Capital Theory

Given the broad range of crowdsourcing applications that we have seen in our review and the fact that organizations can undertake these processes in-house or through intermediaries, the only theoretical model that we are aware of that generalizes the processes and dynamics of all of these specific instantiations is Crowd Capital Theory (CCT). Therefore, we employ CCT ([6, 16], see Figure 1 below), and introduce it in detail to structure our empirical investigation.

### 3.1 Crowd Capital Theory Overview

Crowd Capital is an organizational-level heterogeneous resource generated from IT-mediated crowds. From the perspective of the organization, an IT-mediated crowd can exist inside of an organization, exist external to the organization, or a combination of the latter and the former.

**Figure 1 –Crowd Capital Theory**

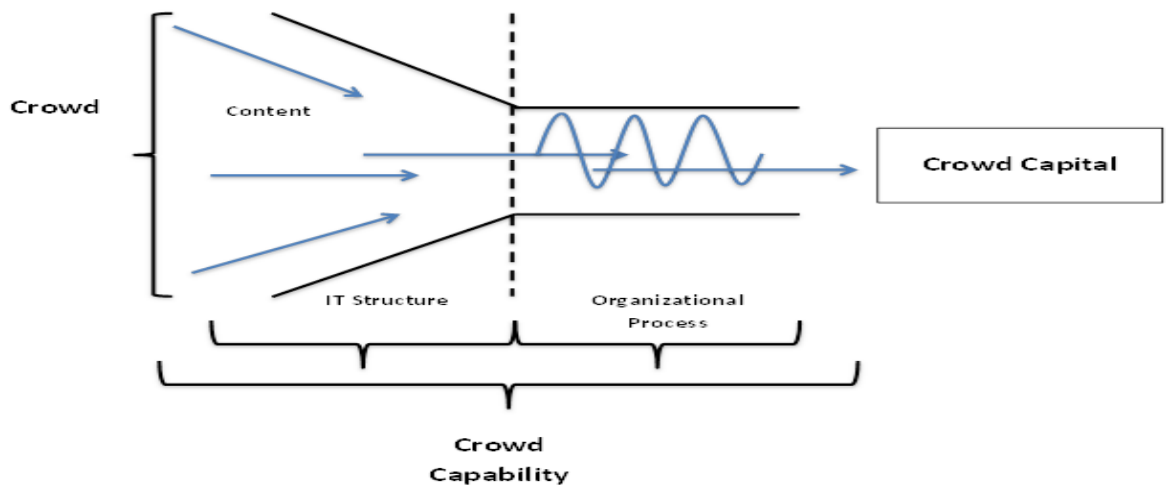

Crowd Capital resource generation is always an IT-mediated phenomenon, and is actuated through an organization's Crowd Capability - an organizational-level capability encompassing the three dimensions of; Content, IT Structure, and internal organizational Processes.

The Content dimension of Crowd Capability defines the form of the content sought from a crowd (e.g. knowledge, information, data, money); the Structure component of Crowd Capability defines the technological means employed by an organization to engage a crowd; and the Process dimension of Crowd Capability refers to the internal procedures that the organization will use to organize, filter, and integrate the incoming crowd-derived contributions.

### 3.2 Dispersed Knowledge

Dispersed knowledge is the antecedent construct of CCT. The existence of dispersed knowledge has been the subject of inquiry in economics for many years, and central to our understanding of dispersed knowledge is the contribution of F.A. Hayek, who in 1945 wrote a seminal work titled 'The Use of Knowledge in Society'. In this work, for which Hayek was eventually awarded the Nobel prize, Hayek describes dispersed knowledge as "…the knowledge of the circumstances…never exists in concentrated or integrated form but solely as the dispersed bits of incomplete and frequently contradictory knowledge which all the separate individuals possess" [17].

Therefore, dispersed knowledge in CCT describes why crowds are useful for organizations to engage. A crowd, comprised of collection(s) of independently-deciding groups or individuals [18, 19], represents a subset of all of the dispersed knowledge available in society at large. And because dispersed knowledge changes moment to moment due to temporal factors, no crowd, let alone any particular group or individual's knowledge is static. Thus, every crowd, even those comprised of the exact same individuals and groups, is always, and everywhere, unique.

### 3.3 Crowd Capability

Crowd Capability is an organizational-level capability that encompasses the Structure, Content, and Process of an organization's engagement with a crowd. The Content dimension represents the form of content sought from a crowd. Well-known forms of content that are currently being sought-out from crowds include micro-tasks [1, 2], ideas and creativity [4, 13, 20], money [21] and technical innovative solutions [9]. The Process dimension of Crowd Capability refers to the internal procedures that the organization will use to organize, filter, and integrate the incoming crowd-derived content contributions. The Structure component of Crowd Capability indicates the technological means employed by an organization to engage a crowd, and crucially, IT structure can be found to exist in either Episodic or Collaborative form, depending on the interface of the IT used to engage a crowd.

With Episodic IT Structures, the members of the crowd never interact with each other individually through the IT. A prime example of this type of IT Structure is Google's reCAPTCHA [22], where Google accumulates significant knowledge resources from a crowd of millions, though it does so, without any need for the crowd members to interact directly with one another through the IT.

On the other hand, Collaborative IT Structures require that crowd members interact with one another through the IT, for resources to be generated. Therefore, in Collaborative IT Structures, social capital must exist (or be created) through the IT for resources to be generated. A prime example of this type of IT structure is Wikipedia, where the crowd members build directly upon each other's contributions through time.

### 3.4 Crowd Capital

Crowd Capital is an organizational-level heterogeneous resource generated from IT-mediated crowds. We label this newly emergent organizational resource as Crowd Capital because it is derived from dispersed knowledge (the crowd), and because it is a key resource (a form of capital) for an organization, that can facilitate productive and economic activity [23]. Like the other forms of capital (social capital, financial capital etc.), Crowd Capital requires investment (for example in Crowd Capability), and potentially leads to literal or figurative dividends, and therefore from our perspective, it is endowed with typical "capital-like" qualities. Further, in respect to Crowd Capital Theory, the Crowd Capital construct is the outcome (or a potential outcome) of engaging IT-mediated crowds.

### 3.5 Hypothesis

As crowdsourcing applications mushroom all around the world, the growing consensus is that resources from people all around the world are available to these companies. An important part of CCT indicates to us that a firm needs to access a crowd of dispersed knowledge before it can generate resources, and after accessing the crowd, the crowd must participate for resources to potentially be generated. In this respect, we wonder how regional location impacts crowd participation? Answering this query is the central thesis of this work, and the results will be a small step in the direction of understanding some differing traits of geographically dispersed crowd participants.

We reason that despite the broad reach of online contest platforms, Crowd Capital participation will still be regionally concentrated. For the purposes of the paper, we are not interested in establishing the causal factors behind regional preferences but rather establishing it in the context of crowdsourcing and CCT, that regional preferences exist and that they are correlated to Crowd Capital participation. Therefore, we propose that:

Ceteris paribus, crowd participants from regions located on the platform's continent of origin, Europe, will be more likely on average to participate in crowdsourcing contests.

## 4. Research Methods

In this section we outline the operationalization of CCT with the data collected from the eYeka platform. We begin by describing the empirical context of the data used, the fit between our research questions and the methods and setting, and then map the operationalizations of the data to the constructs of CCT. From here we describe our data analysis techniques, outlining the impact of geography on Crowd Capital participation.

### 4.1 The Empirical Context

Data were provided to us by eYeka, a leading global crowdsourcing platform based in France which operates a website on which brands and organizations can host creative contests [24, 13]. eYeka hosts crowdsourcing contests on behalf of major companies like Procter & Gamble, Kraft, Coca-Cola, Unilever, Nestle, Danone, and Microsoft. Since the year 2011, all contests that are displayed on eYeka's platform are available at least in English as well as in up to 11 more languages like French, Spanish, Indonesian, Chinese or Russian (a single contest can be translated to up to 12 languages). In early 2014, the eYeka community consisted of about 280,000 members from more than 160 countries. The international brand and vast geographical reach of the platform make it a good setting to study the global geographic distribution of Crowd Capital creation.

To participate, individuals must join the eYeka community (at no cost) by selecting an anonymous username. eYeka collects basic demographic information as self-reported by crowd members (age, gender, skill-level, geographic origin). They can browse through contests on the eYeka website, which are available in English as well as several other languages, and participate in any contest that they feel attracted to without geographical restrictions[1] [25]. Our data contains 1,858,202 observations from submissions to 94 contests of the year 2012. We first assume that all members can see all the contests at all times. We arrive at 1,858,202 contest-participant pairs by excluding all participants who joined the platform after a contest was announced and all participants that have been idle, that is; have not participated in any contests before.

More descriptive statistics about the geographical distribution of participation (indicated as accepted submissions, which are contest submissions judged as relevant by eYeka's community managers) are shown in Table 1 for continents.

**Table 1: Accepted Submissions by Continent**

| Continents | Count | Percent | Cumulative |
|---|---|---|---|
| Europe | 4,413 | 55.64 | 55.64 |
| Asia | 2,183 | 27.52 | 83.17 |
| Latin America | 808 | 10.19 | 93.36 |
| Northern America | 324 | 4.06 | 97.44 |
| Africa | 185 | 2.33 | 99.77 |
| Oceania | 18 | 0.23 | 100.00 |

### 4.2 Mapping the Operationalizations

As outlined in Figure 2 below, we map the operationalizations of CCT to the theoretical model itself. For the purposes of this particular study, we measure dispersed knowledge by partitioning crowd members by continent, and we measure the process construct as our dependent variable, operationalized the number of submissions screened by eYeka.

As outlined in Figure 1, CCT has three major constructs: Dispersed knowledge (antecedent condition), Crowd Capability (comprising of the Content, Structure, and Process dimensions), and Crowd Capital (the resources generated from the crowd). By using eYeka data, we are able to capture variation across the first two constructs of CCT. We operationalize Crowd Capability with the eYeka data, and in respect to the dimensions of Crowd Capability we undertake the following.

---

[1]Exceptions are made for contests with legal restrictions, such as contests for alcohol or tobacco products. Our data do not contain restricted contests.

In respect to IT Structure, eYeka employs an Episodic Structure, where the individual crowd members do not interact with one another through the IS. We therefore hold this constant in our analysis, since it is the same for every participant and every project.

Further, in respect to the submissions from the crowd for each contest, crowd members are not allowed to see the submissions from other crowd members, and thus, each submission is likewise independent in its own right, which we also hold constant in our analysis.

In respect to process dimensions of Crowd Capability, eYeka employs processes on behalf of the client sponsor to filter and vet incoming submissions, in our analysis this is captured by our measures for submission screening, and in this setting it is the dependent variable for crowd participation.

In this particular study we ignore the follow-on step from participation (which is Crowd Capital creation), therein limiting our analysis to participation as the outcome variable of interest.

**Figure 2 - Operationalizations**

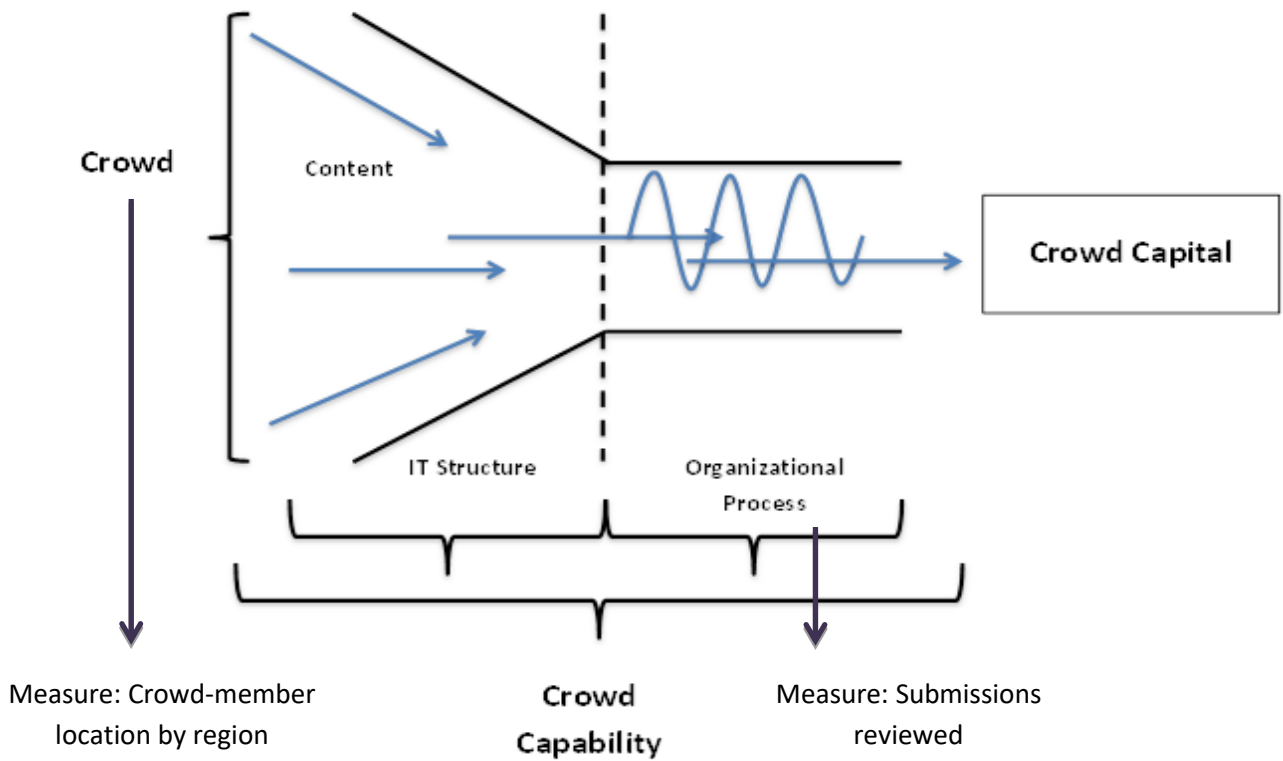

Measure: Crowd-member location by region

Measure: Submissions reviewed

### 4. 3.Data analysis and Methods

We recognize that several different methodologies can be used. We explored the suitability of several econometric techniques—standard heckman, heckman probit—to test our hypothesis. We proceed in similar vein as previous work [26] but modify the techniques to suit our need. In particular, we exploit the fact that eYeka uses a step-by-step process for contest submissions to test for the geographic influences on Crowd Capital Participation.

Participating in contests on eYeka entails several major steps: the first being submitting an entry. After submission, the contest administrator screens the spam and impertinent entries out. The rest are "accepted" into the contests. This is the dependent variable in our study.

To facilitate this analysis, we first matched each of 94 contests to every active member of the eYeka community (at the time, 28,214 individuals), resulting in 2,652,116 contest–participant pairs. We next compared the contest end dates with the dates of the participants' enrollment in eYeka, eliminating instances when the contest ended before the participant joined, resulting in 1,858,202 usable data points remaining. We used STATA's *probit* command to run analyses, clustering at the participant level, because error terms for a participant who entered multiple contests might be correlated[2] following general convention in the literature. We run an estimation of the regression model for submissions using geographic covariates (baseline equation). In terms of geographic covariates, we run two separate sets of regressions for continent and country variables, though here due to space limitations we present only the results for regions.

## 5. Results

We investigate the relative likelihood of crowd members from different continents to submit entries to contests (Crowd Capital participation) to test our proposed hypothesis.

### 5.1 Likelihood to Participate – Partitioned Geographies

In respect to Crowd Capital participation across continents, we find that non-European crowd members are significantly less likely to participate when compared to European crowd members. This supports our hypothesis. See Table 2 for the results of our regression.

Indeed, crowd members from Asia (b=-0.1791, $p<0.001$), Latin America (b=-0.1348, $p<0.001$), North America (b=-0.1297, $p<0.001$) and Oceania (b=-0.5892, $p<0.001$) are significantly less likely to participate in creative crowdsourcing than European crowd members (regression results for Africa are not statistically significant).

---

[2] Multiple participants enter a given contest, but they complete their creative work and submission independently; thus there are unlikely to be strong correlations among the error terms associated with multiple participants' entry into the same contest. Nevertheless, we ran corresponding analyses clustering at the contest and found the same results.

**Table 2: Crowd Capital Participation**

| Crowd Capital Participation - Continent Regressions | |
|---|---|
| Africa | 0.1413 |
| Asia | -0.1791*** |
| Latin America | -0.1348*** |
| North America | -0.1297** |
| Oceania | -0.5892*** |
| Constant | -2.6322*** |
| Observations | 1858202 |
| ** $p<0.01$ *** $p<0.001$ | |

## 6. Discussion

The purpose of this paper was to address an often cited but rarely unpacked aspect of crowds: their geographical distribution and the effect that this has on Crowd Capital participation. While practitioners and academics alike praise the wide-reaching nature of the crowds in crowdsourcing, which allows organizations to tap into a "global talent pool to accelerate innovation" [27] few people have sought to unpack whether this is actually the case.

By analyzing data from a global and open creative crowdsourcing provider, we find that participation in crowdsourcing contests is not equally distributed across the globe. We find that, non-European crowd members are less likely to participate in crowdsourcing (which, does not target any specific audience upfront) compared to European crowd members. In other words, it seems that crowd members from countries closer to the country of the platform's location are more likely to participate in crowdsourcing endeavors.

### 6.2 Theoretical and managerial implications

To our knowledge, our findings are novel in the open innovation and crowdsourcing literature. Indeed, we are not aware of any papers that look at the differences between continents in terms of crowd engagement. While many papers have delved into the effects of contest design on crowdsourcing participation and performance, few papers have sought to unpack the geographical heterogeneity of crowdsourcing [28] and shed light on the, regionally concentrated nature of crowdsourcing. We think that the main reason for this theoretical shortcoming is that most studies have taken place with data from crowdsourcing platforms that are limited in terms crowd diversity [29], making it hard to compare different crowds without changing the empirical setting. In addition to validating the thesis that the commerce and trade is not flat [30, 31] even in fields leveraging pervasive technologies, we contribute to the CCT by providing a first known empirical test of the model.

Managers who want to solicit a global crowd should be aware that even if crowdsourcing platforms promise global participation, global participation remains (statistically-speaking) more local than global. Crowdsourcing is uncertain and draws participation from across the globe [5], but it seems that crowd members shun participating in contests organized by remote platforms. What does this mean for companies or entrepreneurs who want to leverage the workforce or the creativity of the crowd? While the fundamental meritocratic principle of crowdsourcing – the one that says that participation is open to all regardless of their origin and that judgment is based only on the quality of the crowd's output – is not threatened, we see that it's at least challenged. Managers and entrepreneurs should be aware that participation will be significantly more "local" even if outreach is "global", and these facts may materially influence how contest-sponsors structure their contest designs. Similarly, the crowdsourcing platforms should be interested in these findings as the results signal that platforms may not be fully leveraging the diversity that already exists within their crowds.

### 6.3 Limitations and further research

As with any other study, ours is not without limitations. First, the empirical setting of our study is bound to particular crowdsourcing platform with its specificities and processes. By being founded and headquartered in Europe, the eYeka platform represents a specific crowdsourcing community to analyze the global distribution of participation. While being global on the company as well as the community side, making it a good setting to explore the geography of crowdsourcing participation, we feel that our findings could have limited external validity and we look forward to future work that supports or falsifies our findings. We reason that studies that include data from multiple platforms and employ comparable difference-in-differences approach or meta-analytics approach can be of tremendous value for the literature in this respect.

Another limitation of our study is that we focus our analysis on participation and likelihood to submit relevant ideas to creative contests, ignoring success in contests. Villaroel and Reis [32] found that, in internal crowdsourcing initiatives, site marginality— defined as being spatially distant relative to the corporate epicenter— was positively associated with better innovation performance. We feel that we could enrich our findings by going an extra step, looking at the likelihood to win contests across geographies, potentially complementing these findings. We are aware that these limitations can be addressed in subsequent papers, and see many directions for future research.

In research regarding tournament platforms an important factor to consider is community members' motivations and personal values—which are known to vary across geographies— and which have been found to impact participation in open source communities [10]. Future research could complement our participation outcomes with cultural data on the country-level to find out whether personal values or cultural norms are related to participation and performance in crowdsourcing.

In the same vein, we also feel that future research could compare outcomes across countries or continents, spotting potential differences –or absences thereof- of crowd activity in identical settings. For example, in a recent paper that looked at crowdfunding performance, Zheng and colleagues [33] found that an entrepreneur's social network ties and the shared meaning of the project between the entrepreneur and sponsors had the same effects on crowdfunding performance in both China and the U.S.

Furthermore, in addition to geo-specific cultural and institutional variables, contest design variables also play an important role in determining participation [34]. We are in the process of collecting more geo-specific cultural and institutional data to address these research opportunities in future work.

### 7. Conclusion

We started this research with a question regarding the importance of geographic location on participation in IT-mediated crowds. In this work, we were able to document that there are indeed significant geographic co-location effects for Crowd Capital participation in crowdsourcing creative contests.

Our work makes several contributions to the literature on open innovation platforms and creative crowdsourcing. In particular, this is the first work to empirically measure Crowd Capital participation in IT-mediated crowds; and to document that in global IT-mediated phenomenon such as crowdsourcing, participation is significantly affected by the co-location of participants and platform.

However, several avenues for research such as considering the cultural and institutional characteristics of the crowd in these decisions leading to the crowd capital participation still remain open. We hope that international business as well as crowdsourcing scholars will share our enthusiasm regarding this promising and novel intersection of these fields and join forces in exploring it further.